\documentclass[oneside,pdflatex,sn-nature]{sn-jnl}

\usepackage{geometry}
\usepackage{graphicx}
\usepackage{multirow}
\usepackage{amsmath,amssymb,amsfonts}
\usepackage{amsthm}
\usepackage{mathrsfs}
\usepackage[title]{appendix}
\usepackage{xcolor}
\usepackage{textcomp}
\usepackage{manyfoot}
\usepackage{booktabs}
\usepackage{algorithm}
\usepackage{algorithmicx}
\usepackage{algpseudocode}
\usepackage{listings}

\theoremstyle{thmstyleone}

\theoremstyle{thmstyletwo}

\theoremstyle{thmstylethree}

\setcitestyle{super,open={},close={},citesep={,}}
\begin{document}

\title[Article Title]{\textbf{Discovery of unobservable parameters via physical embedding}}

\author[1]{\fnm{Le} \sur{Cheng}}\email{chengle25@nudt.edu.cn}

\author[1]{\fnm{Xiaoran} \sur{Liu}}\email{liuxiaoran10@nudt.edu.cn}

\author[1]{\fnm{Lingjin} \sur{Kong}}\email{konglingjin19@nudt.edu.cn}

\author*[1]{\fnm{Haitao} \sur{Zhao}}\email{haitaozhao@nudt.edu.cn}

\author[1]{\fnm{Jun} \sur{Xiong}}\email{xj8765@nudt.edu.cn}

\author[1]{\fnm{Fanglin} \sur{Gu}}\email{gu.fanglin@nudt.edu.cn}

\author[1]{\fnm{Xiaoying} \sur{Zhang}}\email{zhangxiaoying@nudt.edu.cn}

\author[2]{\fnm{Baoquan} \sur{Ren}}\email{renbq@126.com}

\author[1]{\fnm{Jibo} \sur{Wei}}\email{wjbhw@nudt.edu.cn}

\author[1,2]{\fnm{Hao} \sur{Yin}}\email{yinhao@cashq.ac.cn}

\affil*[1]{\orgdiv{College of Electronic Science and Technology}, \orgname{National University of Defense Technology}, \state{Changsha}, \country{China}}
\affil[2]{\orgname{Academy of Military Science}, \state{Beijing}, \country{China}}

\abstract{Recovering a source signal from indirect measurements often requires estimating latent parameters, such as wireless channel states or MRI coil sensitivities, that cannot be directly observed. Here, we introduce Physics-Embedded Inverse Learning (PEIL), in which a learned estimator predicts these parameters and a fixed, physics-based inverse operator uses them to reconstruct the signal, so that training requires only the source signal as supervision. In systems where multiple parameter combinations can reconstruct the signal equally well, the estimator exploits this freedom to coordinate parameters that compensate for residual modelling errors rather than match ground-truth parameters. In high-mobility wireless communications, PEIL discovers task-optimal configurations that outperform baselines given access to ground-truth parameters, enabling zero-shot generalisation and over 20-fold reduction in training data relative to supervised baselines. To test whether these properties extend across physical domains, we apply PEIL to parallel MRI, where it discovers physically interpretable coil sensitivity maps without calibration scans, yielding reconstructions grounded purely in acquired measurements. These results demonstrate that non-identifiability, conventionally a liability, becomes a resource when the learning objective targets reconstruction quality rather than parameter accuracy.}


\maketitle


\begin{figure*}[t]
\centering
\includegraphics[width=0.99\textwidth]{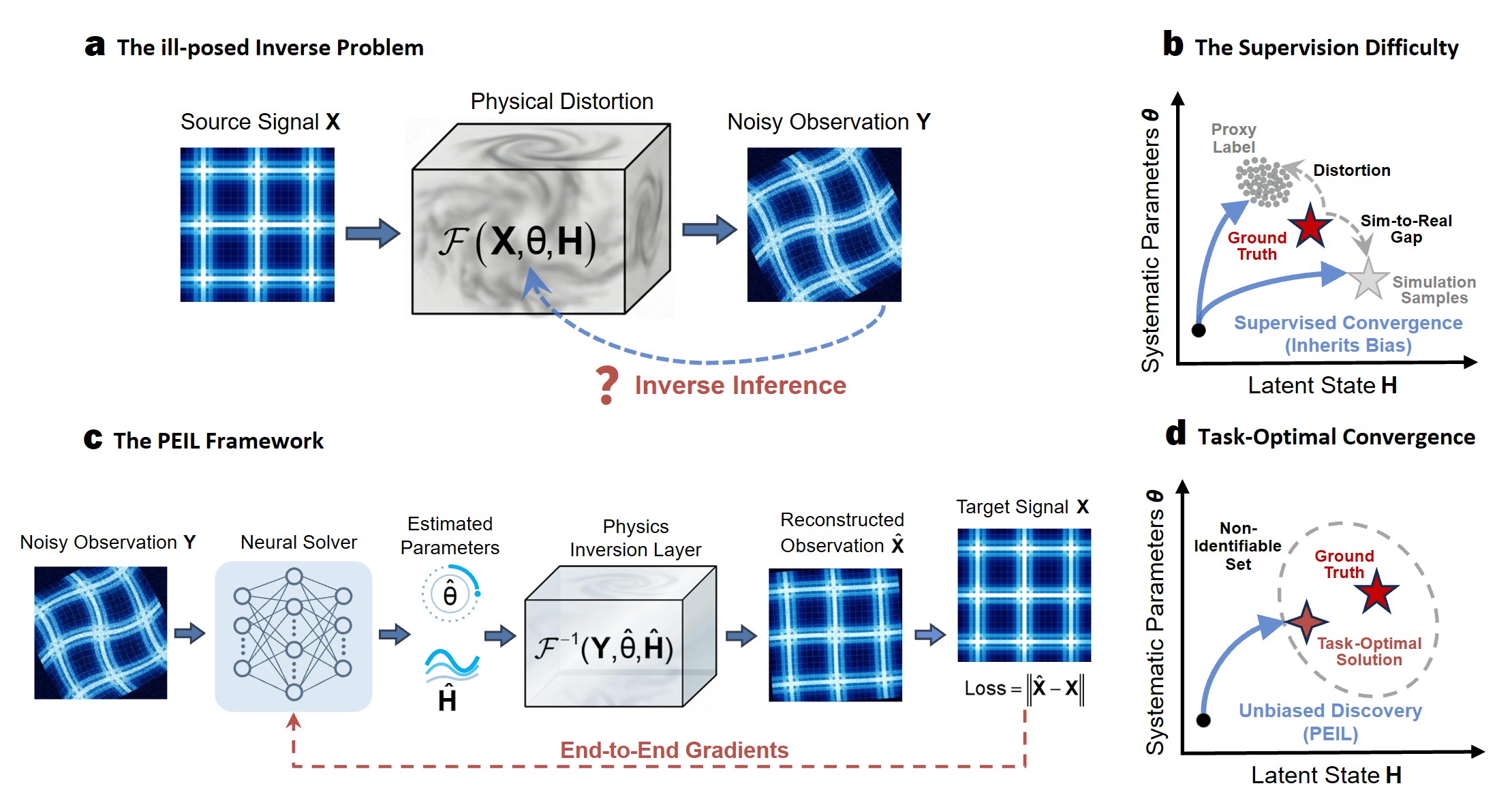}
\caption{\textbf{Problem structure and the PEIL framework.}
\textbf{a,} The ill-posed inverse problem. A source signal $\mathbf{X}$ passes through a forward operator $\mathcal{F}(\mathbf{X}; \boldsymbol{\theta}, \mathbf{H})$, yielding noisy observations $\mathbf{Y}$ governed by unknown systematic parameters $\boldsymbol{\theta}$ and high-dimensional latent parameters $\mathbf{H}$. Jointly recovering $(\mathbf{X}; \boldsymbol{\theta}, \mathbf{H})$ from sparse measurements is ill-posed.
\textbf{b,} The supervision difficulty. True latent parameters are unobservable, forcing supervised approaches to rely on biased proxies from either classical estimation (grey point) or simulation (grey star). The former are corrupted by noise; the latter carry a reality gap. Both bias the learned estimator away from the optimal solution (red star).
\textbf{c,} PEIL. A neural estimator predicts $(\hat{\boldsymbol{\theta}}, \hat{\mathbf{H}})$ to parameterise a fixed, non-learnable inversion layer that reconstructs $\hat{\mathbf{X}}$ from $\mathbf{Y}$ using known physics. Training minimises the reconstruction loss $\|\hat{\mathbf{X}} - \mathbf{X}\|$, enabling gradient flow directly through the analytical operator.
\textbf{d,} Task-optimal convergence. Rather than targeting nominal parameter values, PEIL converges to a task-optimal solution (red four-point star) within the non-identifiable set (grey dashed region), enabling coordinated parameter adjustments that compensate for systematic modelling errors.}
\label{fig1}
\end{figure*}

Many scientific and engineering systems must recover a source signal from indirect, noisy measurements. In high-mobility wireless communications, a receiver must extract the transmitted symbols from observations corrupted by rapid multipath fading and frequency offsets~\cite{you2021towards}. In magnetic resonance imaging (MRI), an anatomical image must be reconstructed from undersampled frequency-domain data acquired through receiver coils with unknown sensitivity profiles~\cite{bib8_Zhu, bib10_Hammernik}. Despite their disparate physical origins, these tasks share a common mathematical structure (Fig.~\ref{fig1}a): a source signal $\mathbf{X}$ passes through a forward operator $\mathcal{F}(\mathbf{X}; \boldsymbol{\theta}, \mathbf{H})$ that couples it with high-dimensional latent parameters $\mathbf{H}$ and systematic parameters $\boldsymbol{\theta}$; only sparse, noisy measurements of the output are observed. Recovering $(\mathbf{X}; \boldsymbol{\theta}, \mathbf{H})$ jointly from such measurements is ill-posed~\cite{bib1_Arridge, bib3_McCann, bib4_Lucas}.

The standard engineering response is explicit calibration: dedicated pilot sequences in wireless systems, or pre-scan acquisitions in MRI. These procedures provide the approximate parameter estimates required to invert the physical model and reconstruct the target signal, but at a direct cost to throughput and operational simplicity. An alternative is to bypass parameter estimation entirely and train a neural network to map observations directly to the target signal~\cite{bib14_LeCun}. This avoids the need for parameter labels, yet discards the known physics of the inverse operator. Without physical structure to constrain the solution space, such models typically require large training sets and generalise poorly to conditions outside the training distribution~\cite{bib28_OShea, bib16_Tobin}. Ideally, one would retain the analytical inverse operator to ensure physically consistent signal recovery, while removing the need for explicit calibration measurements.

A natural middle ground is to keep the analytical inverse operator but train a neural network to estimate the latent parameters, replacing calibration entirely. The difficulty is supervision: ground-truth labels for $(\boldsymbol{\theta}, \mathbf{H})$ are seldom available. In wireless systems, instantaneous channel state information (CSI) is not directly observable~\cite{tse2005fundamentals}; in MRI, exact coil sensitivities lack independent references~\cite{pruessmann1999sense}. In practice, the only parameter labels available are those produced by the calibration procedures the network is meant to replace, or by simulations that carry a reality gap~\cite{bib16_Tobin} (Fig.~\ref{fig1}b).

Even setting this practical obstacle aside, parameter-level supervision has a deeper structural limitation. In systems where multiple latent parameters interact, such as the CSI and frequency offset in orthogonal frequency-division multiplexing (OFDM), or the coil sensitivities and the underlying image in parallel MRI, the mapping from parameters to the reconstructed signal exhibits a fundamental non-identifiability: different parameter combinations can yield equivalent reconstruction quality~\cite{choudhary2014identifiability, li2016identifiability}. Supervised training does not exploit this degeneracy. Since any finite-order physical model inevitably carries residual modelling errors, the nominal parameter values are generally not optimal for reconstruction; coordinated deviations that compensate for model mismatch can yield strictly better signal recovery. Yet supervised training penalises any such deviation, regardless of whether it affects reconstruction quality, and provides no mechanism for parameters to coordinate across the non-identifiable set. What is needed is a learning objective defined not on the parameters themselves, but on the quality of the final reconstruction.

Here we present Physics-Embedded Inverse Learning (PEIL), a decoupled ``estimate-then-solve'' framework that addresses both issues. The approach builds on a structural asymmetry common to these systems: although the complete forward process involves unknown parameters, the underlying physical transformations are well characterised. The Fourier relationship in MRI and the linear channel model in OFDM, for instance, can be inverted analytically and differentiably once the relevant parameters are known. PEIL embeds this analytical inverse operator as a fixed, non-learnable layer within the network (Fig.~\ref{fig1}c). A neural estimator predicts latent parameters $(\hat{\boldsymbol{\theta}}, \hat{\mathbf{H}})$, which are fed into the fixed operator to reconstruct $\hat{\mathbf{X}}$ from $\mathbf{Y}$. Because the operator has no learnable degrees of freedom, reconstruction quality depends entirely on the estimated parameters. Training minimises the reconstruction loss $\|\hat{\mathbf{X}} - \mathbf{X}\|$, leveraging the analytical differentiability of the fixed operator~\cite{bib22_Belbute, bib23_Degrave, bib24_Holl,bib61_gilton2021deep, bib62_pmlr-v119-grazzi20a, bib63_yu2025bilevel} to backpropagate signal-level errors directly into the parameter estimator; the readily available source signal thus serves as implicit supervision for parameter estimation, without requiring ground-truth parameter labels. This resolves the supervision difficulty: the only ground truth required is the target signal itself, not the latent parameters. It also addresses the deeper structural limitation: since the objective is reconstruction quality rather than parameter accuracy, the estimator is free to converge to any point within the non-identifiable set that best serves the final reconstruction (Fig.~\ref{fig1}d), enabling coordinated parameter adjustments that compensate for systematic modelling errors.

Several existing frameworks also combine physics with learned components, but each retains a degree of flexibility that can mask parameter estimation errors. Physics-informed neural networks~\cite{bib20_Raissi, bib21_Karniadakis, bib45_Hemachandra} encode governing equations only as soft loss penalties; deep unrolling methods~\cite{bib25_Monga, bib26_Gregor, bib27_Zhang, bib47_Jacome, bib35_Shlezinger} interleave learnable modules with iterative solvers, allowing the learned components to compensate for inaccurate parameters; and analysis-by-synthesis approaches such as gradSim~\cite{bib68_gradsim} embed a differentiable forward model and optimise in observation space without requiring an explicit inverse. PEIL removes all such flexibility. The fixed inverse operator acts as a hard architectural constraint that concentrates the network's representational capacity on parameter estimation alone: every parameter error directly degrades reconstruction, with no learnable pathway to absorb or mask them. Because improving parameter estimates is the only route available to reduce reconstruction loss, the estimator is driven to exploit structured relationships among the latent parameters, including the coordinated parameter compensation and data efficiency observed in our experiments.

We validate PEIL on two inverse problems that test complementary aspects of the framework. The primary evaluation is high-mobility wireless communications, a regime dominated by rapid non-stationarity and strong parameter coordination~\cite{yang2019deep}. Here, PEIL reduces sample complexity by more than 20-fold and generalises zero-shot across unseen channel profiles and velocities up to three times the training range. In the high signal-to-noise ratio (SNR) regime, it even achieves lower symbol error rates (SER) than an oracle-aided baseline with access to ground-truth parameters. To probe whether these properties extend across physical domains, we apply PEIL to parallel MRI~\cite{pruessmann1999sense}, where the forward operator couples coil sensitivities with the underlying image through a fundamentally different physical mechanism. Without dedicated calibration scans, PEIL estimates coil sensitivity maps that are physically interpretable and sufficient for faithful anatomical reconstruction. Across both settings, the learned parameters deviate from their nominal values yet consistently yield superior or equivalent reconstruction, demonstrating that the framework finds task-optimal solutions within the non-identifiable set rather than recovering a unique nominal parameterisation.

\section*{Results}\label{sec2}

\subsection*{Performance comparison in high-mobility wireless channels}\label{subsecperformance}
 
\begin{figure*}[t!]
\centering
\includegraphics[width=0.99\textwidth]{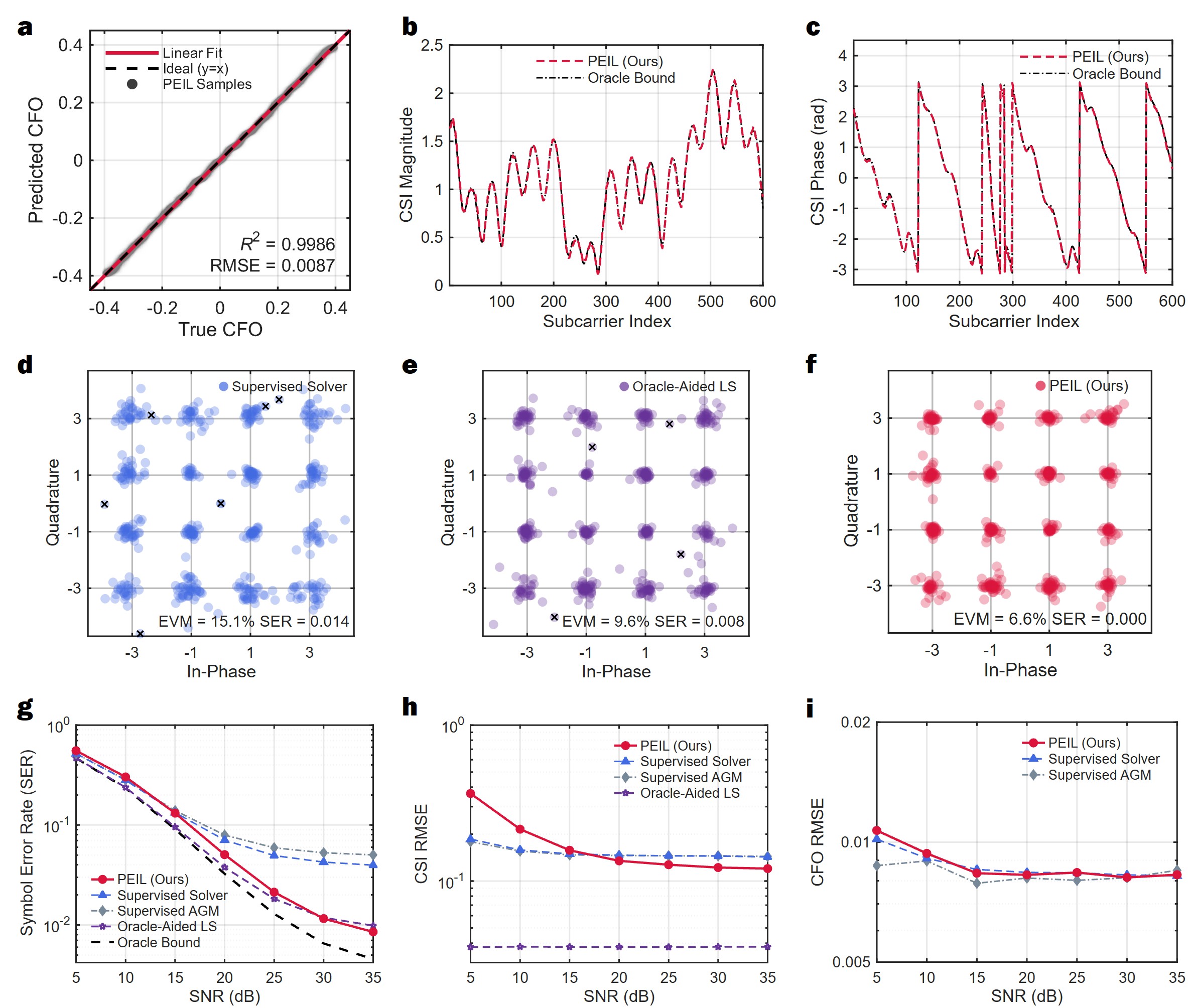}
\caption{\textbf{Wireless signal recovery performance under EVA fading with carrier frequency offset.}
\textbf{a--c,} Implicit recovery of latent parameters. Estimated systematic drift aligns closely with ground truth ($R^2 = 0.9986$; \textbf{a}). Reconstructed CSI profiles capture fine amplitude notches (\textbf{b}) and sharp phase discontinuities (\textbf{c}).
\textbf{d--f,} Target-signal restoration at 30\,dB SNR. Supervised Solver yields dispersed constellations (error vector magnitude (EVM) = 15.1\%, SER = 0.014; \textbf{d}). Oracle-Aided LS tightens the clustering but retains residual distortion (EVM = 9.6\%, SER = 0.008; \textbf{e}). PEIL yields the tightest clustering with minimal errors (EVM = 6.6\%, SER = 0.000; \textbf{f}).
\textbf{g--i,} Performance scaling with SNR. PEIL surpasses Oracle-Aided LS in SER at high SNR regimes (\textbf{g}), while exhibiting slightly higher individual parameter errors (\textbf{h,i}), indicating that the estimator deviates from nominal values to reduce overall reconstruction error.}
\label{fig2}
\end{figure*}
 
We first evaluate PEIL in a highly dynamic wireless system characterised by 3GPP Extended Vehicular A (EVA) fading~\cite{bib52_3GPP} with strong carrier frequency offset (CFO) coupling. Here, the systematic parameter $\boldsymbol{\theta}$ corresponds to the frequency drift introduced by Doppler shift and oscillator mismatch, and the latent parameter $\mathbf{H}$ to the instantaneous CSI. Neither is directly observable, and they interact multiplicatively within the forward operator, creating the joint non-identifiability described above.
 
To benchmark performance, we compare PEIL against theoretical limits and practical baselines. The theoretical tier includes the Oracle Bound~\cite{bib32_Kay} and Oracle-Aided LS~\cite{bib33_Coleri}, an analytical method that uses ground-truth parameters but remains subject to the interpolation error inherent in the fixed inverse operator. For practical comparison, we include Supervised AGM~\cite{bib34_Chen}, a state-of-the-art supervised estimator trained on explicit parameter labels, and Supervised Solver, which shares PEIL's architecture but is trained with parameter labels rather than reconstruction loss, isolating the effect of the training objective alone.
 
Without any parameter supervision, PEIL recovers both latent quantities with high fidelity (Fig.~\ref{fig2}a--c). The estimated CFO drift closely tracks the ground truth ($R^2 = 0.9986$; Fig.~\ref{fig2}a), and the recovered CSI preserves fine structure including deep fading notches in magnitude (Fig.~\ref{fig2}b) and sharp phase discontinuities (Fig.~\ref{fig2}c). Optimising for signal recovery alone is sufficient to disentangle the systematic drift from the CSI, without explicit parameter labels.
 
These recovered parameters, however, are a means, not the end goal: PEIL optimises reconstruction directly. We examine this by comparing signal restoration across methods at 30\,dB SNR (Fig.~\ref{fig2}d--f). Supervised Solver, which shares PEIL's architecture and training data, yields dispersed constellations (Fig.~\ref{fig2}d). Oracle-Aided LS tightens the clustering but retains visible distortion (Fig.~\ref{fig2}e): the fixed inverse operator introduces interpolation errors that ground-truth parameters, which reflect the true channel rather than the operator's approximation, cannot compensate for. PEIL produces the most compact constellation with clear symbol separation (Fig.~\ref{fig2}f), outperforming even Oracle-Aided LS. This result demonstrates that the estimator has converged to parameter configurations that compensate for the operator's inherent approximation errors, yielding reconstructions that ground-truth parameters cannot match. The contrast with Supervised Solver reinforces this conclusion: identical architectures trained on the same data produce vastly different reconstruction quality, isolating the training objective as the decisive factor.
 
Performance scaling with SNR further illuminates the relationship between parameter accuracy and reconstruction quality (Fig.~\ref{fig2}g--i). Below 15\,dB, where the gradient signal from reconstruction loss is weakest, supervised baselines hold a slight advantage. Above 20\,dB, they saturate, while PEIL continues to improve with measurement precision, surpassing Oracle-Aided LS beyond 30\,dB (Fig.~\ref{fig2}g). This is notable because Oracle-Aided LS operates with zero CFO estimation error and uses the true channel response at pilot positions; its residual CSI error (Fig.~\ref{fig2}h) therefore reflects solely the approximation inherent in the fixed inverse operator, an irreducible floor that ground-truth parameter values cannot eliminate. PEIL exhibits higher individual parameter errors than this floor (Fig.~\ref{fig2}h,i), yet achieves lower SER (Fig.~\ref{fig2}g). The parameter deviations are therefore not estimation failures but coordinated adjustments that compensate for the very approximation errors that ground-truth values alone cannot overcome. The next section analyses the quantitative structure of this coordination.

\subsection*{Estimation behaviour under the fixed inverse operator}\label{subsecmechanisms}

\begin{figure*}[t!]
\centering
\includegraphics[width=0.99\textwidth]{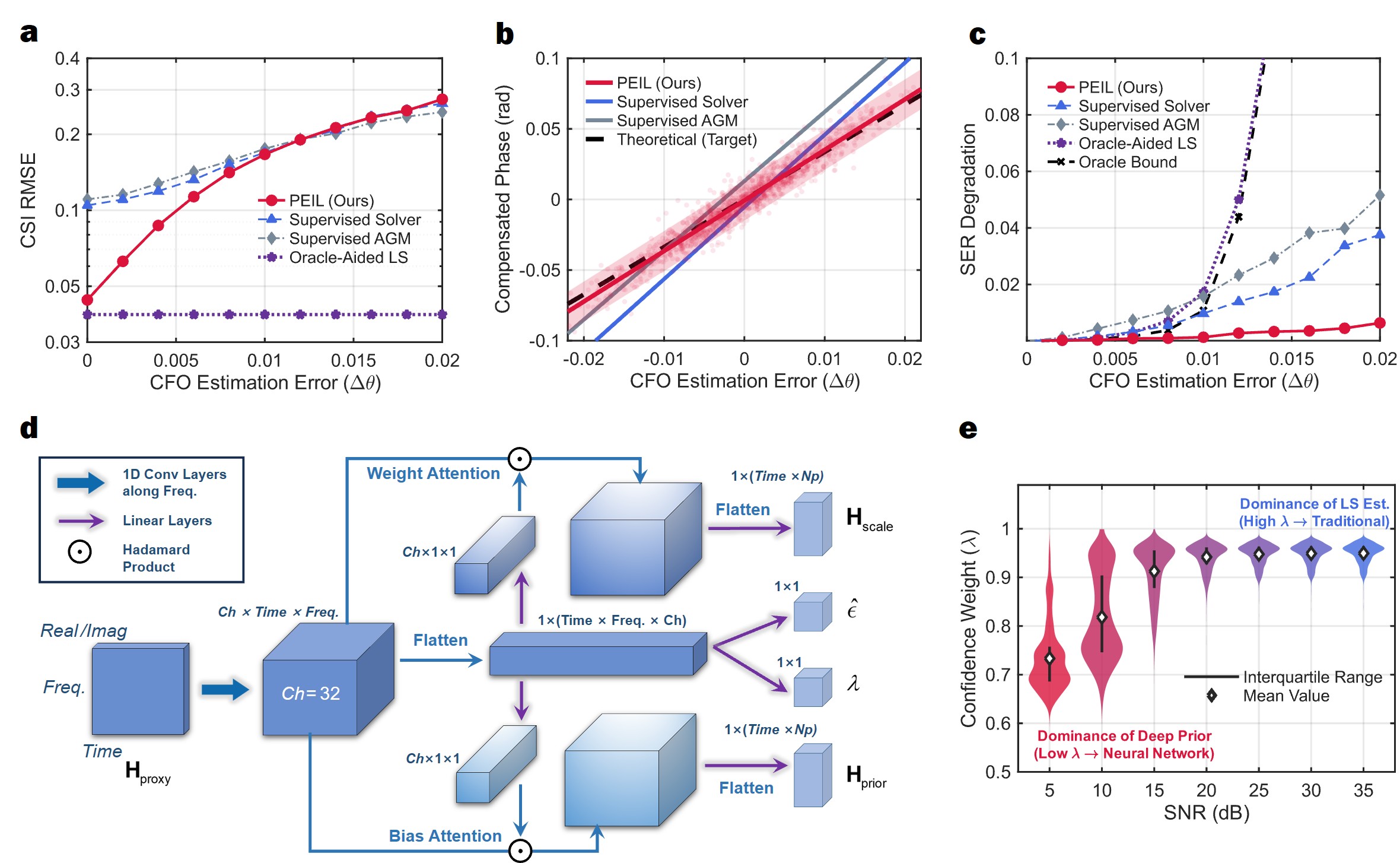}
\caption{\textbf{Learned estimation behaviour under the fixed inverse operator.}
\textbf{a,} Parameter coordination under CFO perturbation. CSI estimation error as a function of CFO estimation error ($\Delta\theta$). Oracle-Aided LS (purple, flat) is unaffected. Both supervised baselines degrade from a higher baseline. PEIL (red) starts lowest but increases progressively, indicating structured coordination between the two parameter estimates.
\textbf{b,} Compensatory phase response. PEIL's channel phase adjustment (red) closely tracks the theoretical compensatory target ($\Delta \phi \approx -\pi \Delta \theta$, black dashed), whereas supervised methods diverge.
\textbf{c,} Reconstruction stability. PEIL maintains near-constant reconstruction quality (red) even under significant CFO perturbation, while all other methods degrade sharply. Oracle-Aided LS, despite having perfect CSI, is equally vulnerable to uncompensated CFO deviation.
\textbf{d,} Gating mechanism. The estimator fuses a coarse physical proxy with a learned deep prior via a learnable confidence gate $\lambda$.
\textbf{e,} Adaptive arbitration. The confidence weight $\lambda$ transitions adaptively with signal quality (violin plots). In noise-dominated regimes (SNR $\le$ 10\,dB), $\lambda$ suppresses the noisy proxy to prioritise the stable deep prior. As precision improves, reliance shifts to the physical observation ($\lambda \to 1$).}
\label{fig3}
\end{figure*}

The previous section showed that PEIL achieves superior reconstruction through parameter configurations that deviate from ground-truth values, and that it remains competitive even in noise-dominated regimes. Here we examine the estimation behaviours underlying each result: a compensatory coordination between parameter estimates that offsets operator approximation errors, and an adaptive gating mechanism that mediates between data-driven and prior-driven estimation as signal quality varies.

We first examine how the estimator responds when one parameter deviates from its nominal value. Figure~\ref{fig3}a plots CSI estimation error as a function of CFO estimation error ($\Delta\theta$). Oracle-Aided LS establishes a lower bound set by the operator's approximation error, as discussed above. Both supervised baselines remain well above this floor even at zero CFO error. PEIL starts near the Oracle floor but its CSI error increases progressively with CFO deviation. Viewed in isolation, this dependence appears detrimental; individual parameter accuracy is worsening. Figure~\ref{fig3}b reveals that it is not. The channel phase adjustment produced by PEIL closely tracks the theoretical phase correction required to compensate for the CFO deviation ($\Delta \phi \approx -\pi \Delta \theta$), whereas supervised methods diverge from this target. The CSI is not degrading randomly; it is rotating in precise opposition to the drift error. 

A separate behaviour addresses the noise-dominated regime. PEIL handles this through a learnable confidence gate $\lambda$ (Fig.~\ref{fig3}d) that mediates between a coarse physical proxy and a learned deep prior. As shown in Fig.~\ref{fig3}e, the gate transitions adaptively with signal quality: in noise-dominated regimes (SNR $\le$ 10\,dB), $\lambda$ remains low, suppressing unreliable observations in favour of the stable prior; as signal quality improves, $\lambda$ approaches unity, shifting reliance to the physical observation. This adaptive arbitration allows PEIL to degrade gracefully at low SNR rather than fail abruptly~\cite{bib37_Kendall}.

Figure~\ref{fig3}c confirms the consequence: PEIL maintains near-constant reconstruction quality even as CFO error grows, while all other methods degrade sharply. Oracle-Aided LS reinforces this point: despite having perfect CSI, an uncompensated CFO deviation alone is sufficient to degrade reconstruction significantly. The contrast underscores that coordination between parameters, not individual parameter accuracy, is the decisive factor. The individual parameter deviations that appear costly in Fig.~\ref{fig3}a cancel in the reconstructed signal, exactly as the non-identifiability of the forward operator permits. Notably, this compensation emerges without explicit design. The fixed inverse operator carries no learnable degrees of freedom, so every uncompensated parameter error propagates directly into reconstruction loss. The resulting gradient signal favours parameter configurations that offset each other and penalises those that do not.


\subsection*{Robust zero-shot generalisation and data efficiency}\label{subsecrobustness}

\begin{figure*}[t!]
\centering
\includegraphics[width=0.96\textwidth]{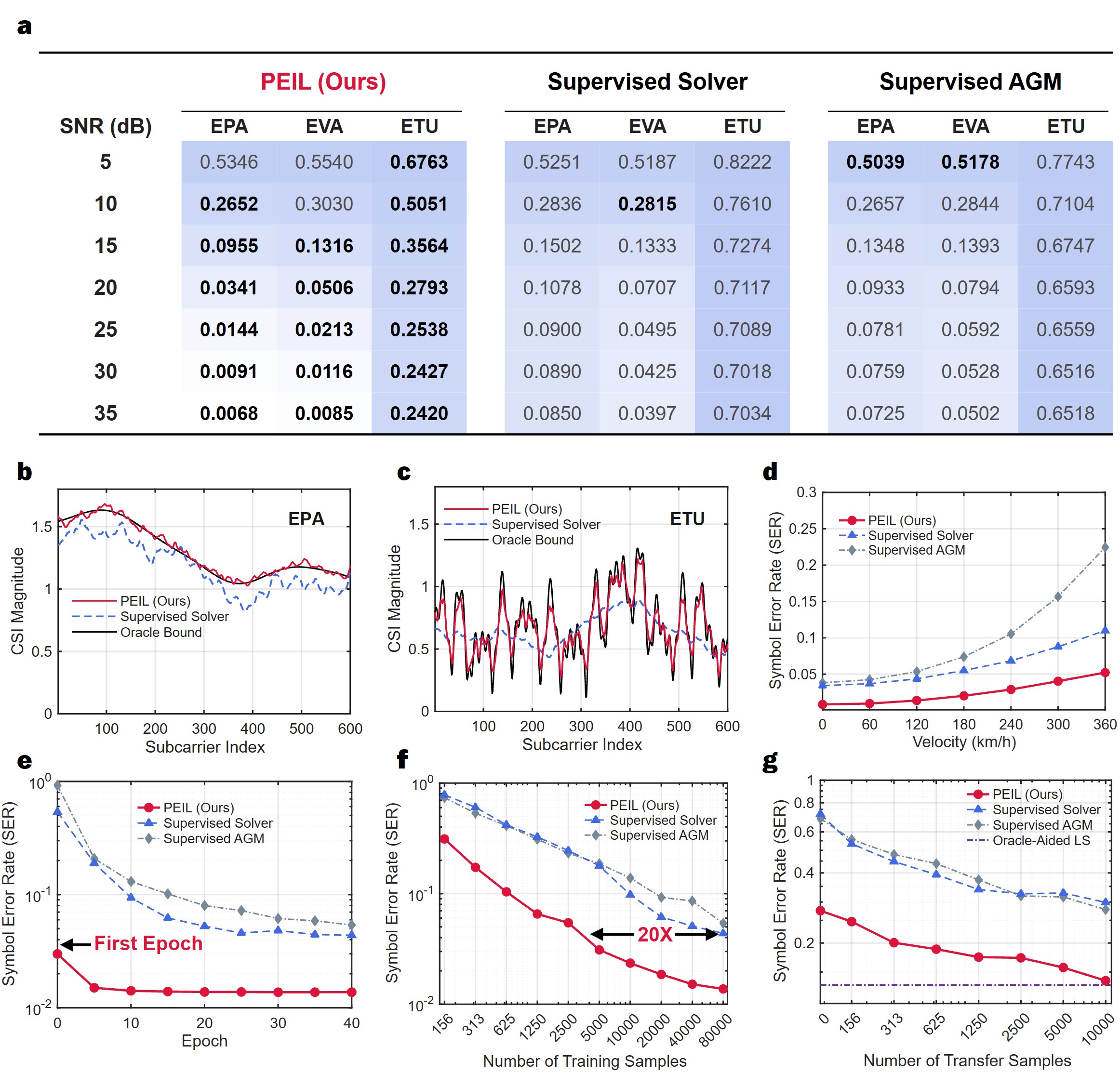}
\caption{\textbf{Generalisation and data efficiency across channel conditions.}
\textbf{a,} Zero-shot structural generalisation. Models trained on medium-complexity dynamics (EVA) are tested on unseen low-complexity (EPA) and high-complexity (ETU) environments. PEIL achieves the lowest SER across the majority of regimes without retraining.
\textbf{b--c,} CSI profiles on unseen regimes. Supervised estimates introduce high-frequency artefacts in simple EPA channels (\textbf{b}) and collapse to mean trends in complex ETU channels (\textbf{c}). PEIL better preserves the spectral structure in both.
\textbf{d,} Kinematic extrapolation. SER versus velocity. PEIL demonstrates stable degradation up to $3\times$ extrapolation (360\,km/h), contrasting with the rapid degradation of supervised baselines.
\textbf{e,} Convergence trajectories. Validation error of PEIL exhibits a sharp drop within the first epoch.
\textbf{f,} Sample complexity. PEIL reduces data requirements by more than 20-fold to reach competitive error floors.
\textbf{g,} Fine-tuning under extreme compound shift (high-mobility ETU). PEIL starts from a functional zero-shot baseline (SER $\approx$ 0.28) and improves rapidly. Supervised methods begin from severe degradation (SER $\ge 0.70$) and require 10,000 adaptation samples to match PEIL's zero-shot performance.}
\label{figgeneralization}
\end{figure*}

The previous sections showed that the fixed inverse operator enables compensatory parameter coordination and superior reconstruction under the training distribution. We now test whether these properties survive when operating conditions shift beyond the training support~\cite{bib39_Shen}.

We train all models exclusively on medium-complexity EVA dynamics at 120\,km/h and evaluate on unseen channel structures and velocities. Generalisation is tested in both directions: toward simpler structure (Extended Pedestrian A, EPA) and toward greater complexity (Extended Typical Urban, ETU). An estimator that has learned the underlying physics rather than the statistical regularities of the training distribution should adapt more easily to a simpler environment and degrade gracefully in a more complex one. Figure~\ref{figgeneralization}b,c reveal that supervised baselines satisfy neither criterion. On the simpler EPA profile, they introduce spurious high-frequency artefacts into their CSI estimates (Fig.~\ref{figgeneralization}b), imposing structure from the training distribution that does not exist in the true channel. On the more complex ETU profile, they collapse to over-smoothed mean profiles that miss deep fading notches (Fig.~\ref{figgeneralization}c). PEIL preserves the spectral structure in both cases. This is reflected quantitatively in Fig.~\ref{figgeneralization}a: on EPA, PEIL achieves the lowest SER at moderate-to-high SNR; on ETU, the contrast is far more pronounced, as the SER of both supervised methods remains largely insensitive to SNR, indicating that their estimates have collapsed and no longer benefit from improved measurement precision, while PEIL continues to scale normally, reducing its SER from 0.68 at 5\,dB to 0.24 at 35\,dB. The pattern extends to kinematic extrapolation (Fig.~\ref{figgeneralization}d): supervised baselines degrade rapidly as velocity increases beyond the training range, while PEIL exhibits stable error scaling up to 360\,km/h, three times the training velocity.

The underlying mechanism is consistent across all three tests. Supervised training must learn to hit a single nominal parameter value in a high-dimensional space shaped by the training distribution; any shift in operating conditions degrades performance. PEIL need only reach the non-identifiable set of parameter combinations that yield valid reconstruction, a target that exists under new operating conditions just as it does under training conditions and that can compensate for systematic errors that cannot be addressed by fitting nominal values alone.

This robustness translates directly into data efficiency (Fig.~\ref{figgeneralization}e--g). PEIL exhibits a sharp drop in validation error within the first epoch (Fig.~\ref{figgeneralization}e), whereas supervised baselines require substantially more iterations to reach comparable performance. Sample complexity is correspondingly reduced: with only 4,000 samples, PEIL attains the performance that supervised baselines require 80,000 to match, a reduction of more than 20-fold (Fig.~\ref{figgeneralization}f). The advantage is most pronounced under severe distribution shift. When fine-tuned on extreme out-of-distribution conditions (360\,km/h ETU), PEIL starts from a functional zero-shot baseline and improves rapidly with few samples (Fig.~\ref{figgeneralization}g). Supervised baselines degrade severely under this distribution shift and require 10,000 adaptation samples to match the performance PEIL achieved before seeing any transfer data. With continued fine-tuning, PEIL approaches the Oracle-Aided LS baseline, demonstrating strong transfer capability even in conditions far beyond the training distribution.


\subsection*{Application to blind calibration in parallel MRI}\label{subsecmri}

To probe whether the properties observed in wireless communications extend across physical domains, we apply PEIL to parallel imaging MRI using the fastMRI knee dataset~\cite{bib57_knoll2020fastmri}. Here, the latent parameters correspond to coil sensitivity maps that traditionally require dedicated calibration scans, and the fixed inverse operator uses these maps to reconstruct the anatomical image from undersampled k-space data. The central question is whether the estimator, trained solely on reconstruction loss through the fixed inverse operator, discovers coil sensitivity maps that are physically interpretable and consistent across acquisition conditions. We therefore structure the comparison around two baselines, each isolating a different factor. Classical sensitivity encoding (SENSE) uses calibration-derived sensitivity maps within the same physical inverse operator, so that performance differences reflect the quality of the estimated maps. A supervised U-Net with equal or greater capacity is trained to output the reconstructed image directly, bypassing the inverse operator, so that any performance gap can be attributed to the embedded physical inversion rather than network capacity.

\begin{figure*}[t!]
\centering
\includegraphics[width=0.99\textwidth]{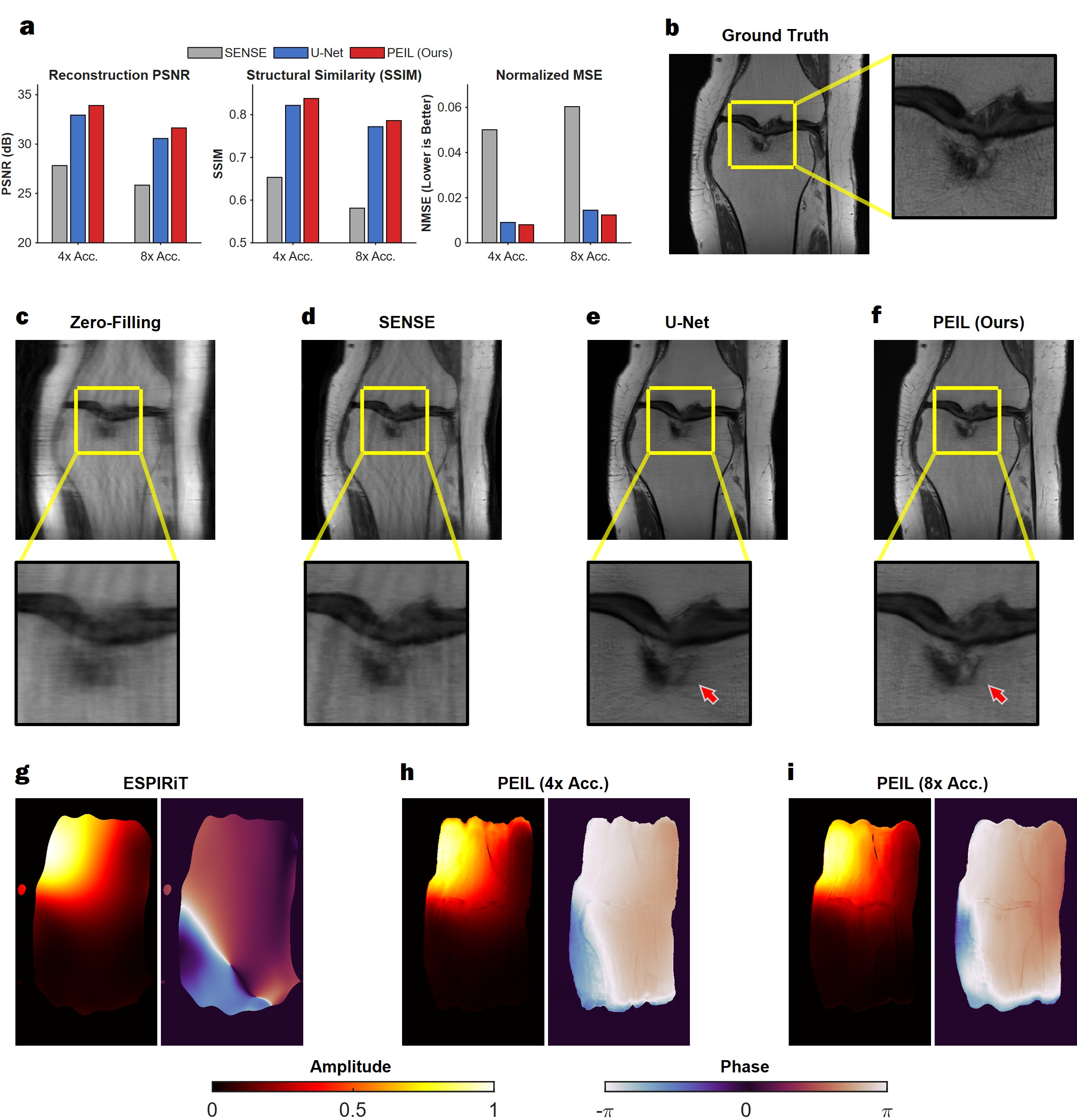}
\caption{\textbf{Blind calibration and image reconstruction in parallel MRI.}
\textbf{a,} Quantitative fidelity on the fastMRI knee dataset. PEIL (red) achieves lower error rates than the supervised U-Net baseline (blue) and classical SENSE reconstruction (grey), particularly at higher acceleration factors ($8\times$).
\textbf{b--f,} Visual reconstruction at $8\times$ acceleration. Zero-filling (\textbf{c}) and SENSE (\textbf{d}) suffer from noise amplification; the supervised U-Net (\textbf{e}) exhibits texture smoothing. PEIL (\textbf{f}) restores sharp meniscal boundaries and fine trabecular textures (red arrows), more closely resembling the fully-sampled reference (\textbf{b}).
\textbf{g--i,} Coil sensitivity map estimation. Comparison of maps (magnitude and phase) estimated by ESPIRiT (calibration reference, \textbf{g}) versus those estimated by PEIL under $4\times$ (\textbf{h}) and $8\times$ (\textbf{i}) acceleration. Despite distinct aliasing patterns, PEIL recovers consistent phase profiles, indicating that the estimator disentangles stable hardware properties from observation-dependent sampling artefacts.}
\label{figmri}
\end{figure*}

We first assess reconstruction quality (Fig.~\ref{figmri}a--f). PEIL outperforms both classical SENSE reconstruction and a supervised U-Net baseline, with the performance gap widening at $8\times$ acceleration (Fig.~\ref{figmri}a). Visually, SENSE suffers from noise amplification and the supervised U-Net exhibits texture smoothing, whereas PEIL better preserves sharp meniscal boundaries and fine trabecular bone textures (Fig.~\ref{figmri}b--f).

To understand what underlies this advantage, we examine the coil sensitivity maps that PEIL discovers (Fig.~\ref{figmri}g--i). Since coil sensitivities are hardware properties, they should remain constant regardless of the undersampling rate. Figure~\ref{figmri}g--i shows sensitivity maps from the ESPIRiT~\cite{bib58_uecker2014espirit} calibration reference alongside those estimated by PEIL under $4\times$ and $8\times$ acceleration. Although different acceleration factors introduce different aliasing artefacts into the input data, the estimated sensitivity profiles remain consistent: magnitude maps closely match the calibration reference, and phase profiles maintain structural coherence across acceleration factors. The estimator, trained through the fixed inverse operator, disentangles stable hardware parameters from transient sampling artefacts, just as the wireless estimator disentangles the systematic drift from the CSI. This consistency indicates that the estimator converges to physically meaningful sensitivity maps rather than compensating for inaccurate maps through the network's predicted image. Because reconstruction is performed by the fixed inverse operator parameterised by these maps, the recovered anatomy is grounded in the acquired measurements rather than hallucinated by a learned prior.

\section*{Discussion}\label{sec3}

We have presented PEIL, a framework that embeds a fixed analytical inverse operator as a non-learnable layer within a neural network. This design resolves two issues identified in the introduction. The first is practical: by minimising reconstruction loss rather than parameter error, PEIL shifts supervision from inaccessible parameter labels to readily available signal-level ground truth. The second is structural: because the objective is reconstruction quality, the estimator is free to exploit the non-identifiability of the forward operator rather than being constrained to a single nominal parameterisation. In wireless channels, coordination between parameters proves more important than individual parameter accuracy: the estimator discovers configurations that compensate for the operator's inherent approximation errors, achieving reconstructions that ground-truth values cannot match, and generalising zero-shot across channel environments by capturing the underlying physics rather than the statistical regularities of the training distribution. In MRI, physically consistent coil maps yield reconstructions grounded in acquired measurements rather than hallucinated by a learned prior. All these results follow from the same principle: non-identifiability, conventionally regarded as a nuisance that complicates estimation, becomes a resource when the learning objective is aligned with the final reconstruction.

The fixed inverse operator distinguishes PEIL from related approaches in a way that has practical consequences. Physics-informed neural networks~\cite{bib20_Raissi, bib21_Karniadakis} enforce physics through soft loss penalties, which requires balancing hyperparameters between data fidelity and physical terms; the balance is problem-specific and often fragile. Deep unrolling methods~\cite{bib25_Monga, bib26_Gregor, bib27_Zhang} and bilevel optimisation frameworks~\cite{bib61_gilton2021deep, bib62_pmlr-v119-grazzi20a, bib63_yu2025bilevel} employ learnable solver components that can absorb estimation errors through adaptive correction, potentially masking physically inconsistent parameter estimates. PEIL removes this possibility: because the inverse operator has no learnable degrees of freedom, every parameter error directly degrades reconstruction, and the training signal is unambiguous. This strictness is both a strength and a constraint. It produces the robust generalisation and compensation behaviours observed in our experiments, but it also limits applicability to problems where an analytical or efficiently differentiable inverse operator is available. In comparison, purely data-driven self-supervised methods~\cite{bib43_yaman2020self} require no physical model at all, offering broader applicability at the cost of forgoing the inductive bias that enables PEIL's data efficiency and out-of-distribution robustness.

From an engineering perspective, PEIL addresses a practical bottleneck in deploying learned models for physical systems. In wireless communications, the 20-fold reduction in sample complexity translates directly to reduced pilot overhead and shorter calibration periods, facilitating rapid deployment and adaptation to changing propagation environments. The zero-shot generalisation across channel profiles and velocities up to $3\times$ the training range suggests that a single trained model could serve across diverse operating conditions without retraining. In MRI, eliminating dedicated calibration scans shortens examination time and simplifies clinical workflows. More broadly, the estimate-then-solve structure is applicable wherever an analytical inverse operator is available and the forward process couples the source signal with unobservable latent parameters. Optical channel equalisation, acoustic source localisation, and sensor array calibration share this structure and could benefit from the same design.

The approach has limitations that define its scope. It requires an analytical or efficiently differentiable inverse operator; highly nonlinear regimes lacking closed-form inverses would require alternative formulations. Embedding iterative solvers such as conjugate gradient iterations incurs higher training costs than pure feedforward baselines, although inference cost is comparable. The blind disentanglement of latent parameters from the target signal is inherently ill-posed and may require domain-specific regularisation to resolve ambiguities; in our MRI experiments, for instance, we introduced a double-well potential to address scaling ambiguity between coil sensitivities and the reconstructed image. The adaptive arbitration behaviour observed in wireless channels, where the estimator learns to balance learned priors against noisy observations, emerged without explicit design; whether analogous behaviours arise in other domains remains an open question. A natural direction for future work is integrating the fixed inverse operator constraint into more expressive architectures, such as unrolled networks~\cite{bib64_sriram2020end}, to combine the generalisation benefits of hard physical constraints with the representational power of learned signal-domain priors, while strictly preserving the non-learnable nature of the physical inversion step.

\section*{Methods}\label{sec4}

\subsection*{Problem formulation}\label{subsecmethod_formulation}

The goal is to reconstruct a target signal $\mathbf{X} \in \mathbb{C}^{N}$ from sparse, noisy observations $\mathbf{Y} \in \mathbb{C}^{M}$ ($M \ll N$)~\cite{bib51_Donoho}. The forward process is governed by a domain-specific physical operator $\mathcal{F}$:
\begin{equation}
    \mathbf{Y} = \mathcal{P}_{\Omega} \left( \mathcal{F}(\mathbf{X}; \boldsymbol{\theta}, \mathbf{H}) \right) + \mathbf{n},
\end{equation}
where $\mathcal{P}_{\Omega}$ denotes the sampling mask and $\mathbf{n}$ is additive noise. The latent parameters $(\boldsymbol{\theta}, \mathbf{H})$ are unknown: in wireless systems, $\boldsymbol{\theta}$ is the systematic CFO and $\mathbf{H}$ the instantaneous channel state; in parallel MRI, $\mathbf{H}$ corresponds to the coil sensitivity maps $\mathbf{S}$ with no systematic parameter. Recovery requires jointly estimating the latent parameters and inverting the physical operator, without ground-truth parameter supervision. This task mirrors blind deconvolution and calibration-free reconstruction~\cite{bib53_Gull, choudhary2014identifiability}, guided here by known physical laws rather than purely statistical priors.

\subsection*{Physics-embedded neural solver}\label{subsecmethod_solver}

PEIL follows a strict estimate-then-solve paradigm: a neural estimator predicts the latent parameters from a coarse proxy $\mathbf{V}_{\text{proxy}} = \mathcal{F}^{\dagger}(\mathbf{Y})$, obtained by applying a deterministic adjoint or pseudo-inverse to the sparse observations, which are then fed into a fixed, non-learnable inverse operator that reconstructs the target signal.

\noindent\textbf{Neural estimator.}
In wireless systems, $\mathbf{V}_{\text{proxy}}$ is the zero-forcing estimate at pilot locations, zero-filled elsewhere. The estimator employs a lightweight 1D convolutional network along the frequency dimension and predicts a dense channel prior $\mathbf{H}_{\text{prior}}$, a multiplicative scaling factor $\mathbf{H}_{\text{scale}}$, the scalar CFO $\hat{\theta}$, and a confidence gate $\lambda \in [0, 1]$. In MRI, $\mathbf{V}_{\text{proxy}}$ comprises zero-filled inverse-FFT reconstructions from all coils. The estimator uses a U-Net backbone~\cite{bib60_ronneberger2015u} and predicts the complex-valued coil sensitivity maps $\hat{\mathbf{S}}$, a de-aliased image prior $\mathbf{X}_{\text{prior}}$, and a regularisation weight $\lambda_{CG}$.

\noindent\textbf{Fixed inverse operator: wireless.}
The operator first compensates for the estimated drift via phase rotation, $\tilde{\mathbf{Y}} = \mathbf{Y} \odot e^{-j 2\pi\hat{\theta}\,\mathbf{t}}$. The dense CSI is then recovered by fusing the sparse proxy and the learned prior through a fixed Gaussian interpolation kernel $\mathcal{K}$:
\begin{equation}
    \hat{\mathbf{H}} = \mathcal{K} \left(\lambda\mathbf{H}_{\text{proxy}} \odot \mathbf{H}_{\text{scale}} + (1 - \lambda)\mathbf{H}_{\text{prior}} \right),
\end{equation}
where the kernel width is calibrated to the pilot density. The target signal is recovered by coherent equalisation: $\hat{\mathbf{X}} = \tilde{\mathbf{Y}} \oslash \hat{\mathbf{H}}$.

\noindent\textbf{Fixed inverse operator: MRI.}
The inversion solves a Tikhonov-regularised problem:
\begin{equation}
\begin{aligned}
\hat{\mathbf{X}} = \mathop{\arg\min}_{\mathbf{X}} \bigl( & \|\mathbf{Y} - \mathcal{P}_{\Omega}(\mathbf{F}(\hat{\mathbf{S}} \odot \mathbf{X}))\|_2^2 \\
& + \lambda_{CG} \|\mathbf{X} - \mathbf{X}_{\text{prior}}\|_2^2 \bigr),
\end{aligned}
\end{equation}
where $\mathbf{F}$ is the 2D FFT. The neural estimator parameterises $\hat{\mathbf{S}}$, $\mathbf{X}_{\text{prior}}$, and $\lambda_{CG}$, but the solver itself has no learnable parameters. In MRI, $\mathbf{X}_{\text{prior}}$ additionally serves as a learned image prior that regularises the severely underdetermined reconstruction at high acceleration factors. We unroll the conjugate gradient algorithm~\cite{bib59_hestenes1952methods} for 7 iterations, retaining differentiability for end-to-end training.

\subsection*{Training}\label{subsectraining}

Although the latent parameters $(\boldsymbol{\theta}, \mathbf{H})$ are unknown, the target signal $\mathbf{X}_{\text{gt}}$ is available during training. The model is trained end-to-end by minimising:
\begin{equation}
    \mathcal{L} = \|\hat{\mathbf{X}} - \mathbf{X}_{\text{gt}}\|_1 + \gamma \mathcal{L}_{\text{phys}}.
\end{equation}
For complex-valued signals, the $\ell_1$ reconstruction loss is computed as $\|\mathbf{z}\|_1 = \sum_k |z_k|$, where $|\cdot|$ denotes the complex modulus. For wireless systems, $\gamma = 0$ and the fixed kernel serves as the sole architectural constraint. For MRI, $\gamma = 0.1$ and
\begin{equation}
\mathcal{L}_{\text{phys}} = \underbrace{\|\mathbf{Y} - \mathcal{P}_{\Omega}(\mathbf{F}(\hat{\mathbf{S}} \odot \mathbf{X}_{\text{gt}}))\|_2^2}_{\text{multi-coil consistency}} + \beta \underbrace{\operatorname{mean}\!\left[(\hat{\mathbf{S}}_{\text{rss}} (\hat{\mathbf{S}}_{\text{rss}} - 1))^2\right]}_{\text{double-well potential}}
\end{equation}
with $\beta = 5$, and $\hat{\mathbf{S}}_{\text{rss}}$ denotes the root-sum-of-squares magnitude of the estimated coil sensitivities. Both the reconstruction loss and the multi-coil consistency term~\cite{bib67_DJsense} supervise at the signal level: the former penalises errors in the reconstructed image, the latter enforces agreement between the estimated coil maps and the acquired observations. Neither term requires ground-truth coil sensitivity labels. The double-well potential encourages the root-sum-of-squares sensitivity magnitude to approach either 0 (background) or 1 (region of interest), resolving the scaling ambiguity inherent in blind calibration.

\subsection*{Experimental details}\label{subsecimpl}
For wireless systems, we simulate a 3GPP LTE-based OFDM downlink (2.6\,GHz carrier, 10\,MHz bandwidth) under the EVA fading model~\cite{bib52_3GPP} with random CFO. Oracle-Aided LS uses ground-truth pilots interpolated via the same Gaussian kernel as PEIL to ensure a fair comparison. For MRI, we use the fastMRI knee dataset~\cite{bib57_knoll2020fastmri} with Cartesian undersampling at $4\times$ and $8\times$ acceleration. Both systems are trained with the Adam optimiser~\cite{bib49_Adam}.

\section*{Data availability}
The wireless communication data generated in this study are based on standard 3GPP LTE-FDD simulations using EPA, EVA, and ETU channel models as detailed in the Methods section. Source data supporting the findings of this study are available from the corresponding author upon reasonable request. The MRI data used in this study are derived from the publicly available fastMRI knee dataset, which can be accessed at \url{https://fastmri.med.nyu.edu/} upon registration.

\section*{Funding}
This work was supported by the Key Program of the Joint Funds of the National Natural Science Foundation of China (Grant No. U2441226), the General Program of National Natural Science Foundation of China (Grant No. 62101569 and 62371462), and the Innovation Research Foundation of the College of Electronic Science and Technology at National University of Defense Technology.

\section*{Author Contributions}
L.C. conceived the idea and designed the methodology. L.C., L.K., and H.Z. developed the software code and performed the validation experiments. L.C., F.G., and X.Z. conducted the formal analysis and theoretical derivations. L.C. and X.L. were responsible for data curation and visualization. L.C. wrote the original draft. H.Z., B.R. and J.W. handled the review and editing of the manuscript. H.Z. and J.X. supervised the project. H.Z., J.W. and H.Y. acquired the resources and funding. All authors contributed to the interpretation of the results and approved the final version.

\section*{Competing interests}
The authors declare no competing interests.

\bigskip
\bigskip

\bibliography{sn-bibliography}

\end{document}